# A three-state prediction of single point mutations on protein stability changes

Emidio Capriotti[1], Piero Fariselli[2], Ivan Rossi[2,3] and Rita Casadio[2] [§]

[1]Structural Genomics Unit, Bioinformatics Department, Centro de Investigación Príncipe Felipe (CIPF), Valencia, Spain.
[2] Laboratory of Biocomputing, CIRB/Department of Biology, University of Bologna, via Irnerio 42, 40126 Bologna, Italy
[3] BioDec Srl, via Calzavecchio 20/2, Casalecchio di Reno Bologna, Italy

[§]Corresponding author. Email addresses: EC: ecapriotti@cipf.es  PF: piero.fariselli@unibo.it   IR: ivan@biodec.com  RC: casadio@alma.unibo.it

## Abstract

### Background

A basic question of protein structural studies is to which extent mutations affect the stability. This question may be addressed starting from sequence and/or from structure. In proteomics and genomics studies prediction of protein stability free energy change ($\Delta\Delta G$) upon single point mutation may also help the annotation process. The experimental $\Delta\Delta G$ values are affected by uncertainty as measured by standard deviations. Most of the $\Delta\Delta G$ values are nearly zero (about 32% of the $\Delta\Delta G$ data set ranges from -0.5 to 0.5 Kcal/mol) and both the value and sign of $\Delta\Delta G$ may be either positive or negative for the same mutation blurring the relationship among mutations and expected $\Delta\Delta G$ value. In order to overcome this problem we describe a new predictor that discriminates between 3 mutation classes: destabilizing mutations ($\Delta\Delta G<-0.5$ Kcal/mol), stabilizing mutations ($\Delta\Delta G>0.5$ Kcal/mol) and neutral mutations ($-0.5\leq\Delta\Delta G\leq 0.5$ Kcal/mol).

### Results

In this paper a support vector machine starting from the protein sequence or structure discriminates between stabilizing, destabilizing and neutral mutations. We rank all the possible substitutions according to a three state classification system and show that the overall accuracy of our predictor is as high as 52% when performed starting from sequence information and 58% when the protein structure is available, with a mean value correlation coefficient of 0.30 and 0.39, respectively. These values are about 20 points per cent higher than those of a random predictor.

### Conclusions

Our method improves the quality of the prediction of the free energy change due to single point protein mutations by adopting a hypothesis of thermodynamic reversibility of the existing experimental data.  By this we both recast the thermodynamic symmetry of the problem and balance the distribution of the available experimental measurements of free energy changes. This eliminates possible overestimations of the previously described methods trained on an unbalanced data set comprising a number of destabilizing mutations  higher than that of stabilising ones.

# Introduction

The measure of the protein stability change upon single point mutations is a thermodynamic quantity whose accurate prediction is a key problem of Structural Bioinformatics. In the last years a significant number of different methods are been developed to predict the stability free energy changes ($\Delta\Delta G$) in protein when one residue is mutated. Some methods developed different energy functions, suited to compute the stability free energy [1-11], while other machine learning approaches [12-15]. The introduction of machine learning approaches follows the increasing number of experimental thermodynamic data and their availability in the ProTherm database [16]. However, these automatic methods suffer from the fact that experimental data are affected by errors. When the value of the free energy change is close to 0 and the associated error is considered, for one single measure the sign of $\Delta\Delta G$ can change from decreasing to increasing and vice versa. Another problem is that the training data are intrinsically non symmetric and unbalanced, with destabilizing mutations outnumbering stabilizing ones. This can bias training and testing, effecting the final statistical performance of the predictors at hand.

In this paper we describe a possible solution to the above-mentioned problems and implement a new predictor able to discriminate between 3 classes (destabilising, neutral and stabilising mutations). The new implementation predicts the free energy changes starting for the protein structure or from the protein sequence with an improved scoring efficiency with respect to our previous implementations that routinely discriminate only two putative classes (destabilising and stabilising mutations). Our present method provides therefore a better discrimination of single mutated residues that may have negligible effects on protein stability.

# Material and Methods

### The protein databases

The databases used in this work are derived from the release (September 2005) of the Thermodynamic Database for Proteins and Mutants ProTherm [16]. We select our initial set imposing the following constrains:
a) the $\Delta\Delta G$ value was extrapolated from experimental data and reported in the data base;
b) the data are relative to single mutations;
c) the data are obtained from reversible experiments

After this procedure we obtain a larger data set comprising 1681 different single point mutations and related experimental data for 58 different proteins. In Figure 1 we report the distribution of the $\Delta\Delta G$ values. From the latter by selecting only 55 proteins known with atomic resolution we have a subset of 1634 mutations. Adopting a criterion of thermodynamic reversibility for each mutation, we double all the thermodynamic data. Finally, we end up with 3362 mutations for the set containing protein sequences (DBSEQ) and 3268 mutations for the subset of proteins known with atomic resolution (DB3D). According to experimental $\Delta\Delta G$ value each mutation is grouped into one of the following three classes:
   i) destabilizing mutation, when $\Delta\Delta G < -0.5$ Kcal/mol;
   ii) stabilizing mutation when $\Delta\Delta G > 0.5$ Kcal/mol;
   iii) neutral mutations when $-0.5 \leq \Delta\Delta G \leq 0.5$ Kcal/mol.

The choice of |0.5| Kcal/mol as a threshold value for ΔΔG classification provides a balanced datasets and is also a limiting value of standard errors reported in experimental works.

In order to test the performance of our method another database was generated (using the selection rules a and b listed above) from the current version of ProTherm (April 07). Moreover, to avoid the introduction of mutations that share similarity with those of the training set, we eliminated from the new database the mutations that occur in sequence positions just considered in the training sets. Finally we obtain a dataset of 34 proteins with 499 mutations. Considering the hypothesis of thermodynamic reversibility and the previous classification rules we have a dataset of 998 mutations (NewDB) in proteins with known 3D structure.

**The thermodynamic assumption**

A possible way to improve a classification task is to try to insert more information in the input code and simultaneously try to refine the quality of the discriminated features. In order to meet this requirement here we implement a new predictor able to discriminate between 3 possible classes, namely: i) destabilizing mutations, which are characterized by a ΔΔG<-0.5 Kcal/mol; ii) stabilizing mutations when ΔΔG>0.5 Kcal/mol and iii) neutral mutations when the -0.5 ≤ ΔΔG ≤ 0.5 Kcal/mol. The problem of the asymmetric abundance of the three classes is addressed assuming that from the point of view of basic thermodynamics a protein and its mutated form should be endowed with the same free energy change, irrespectively of the reference protein (native or mutated). If this is so, we can assume that the module of free energy change is the same in going from one molecule to the other and that what changes is only the ΔΔG sign. By this, given a free energy value derived experimentally from a protein mutation, we can take advantage of the previous statement and use the reverse mutation (namely the mutation that transforms back the mutant into the original protein) by considering the value of the experimental measure with the opposite sign (-ΔΔG). The number of the available data in the training set doubles and as a nice side-effect we also balance the training dataset overcoming the problem of the skewness of the experimental data.

Obviously one may pose the question if this observation that is formally correct from the thermodynamic point of view is also applicable to the protein structure and sequence. Providing that we adopt the approximation that local environment plays a dominant role (spatial or sequence-neighbour only) this approach is formally correct. If we start from the protein sequence the formal statement is correct. When the structural environment is taken into account, the local approximation may break down, and spatial rearrangement may happen. In this case using only one structure to compute the local environment for both the mutation and its reverse may be inaccurate. However, all predictive approaches developed so far, including those based on energy functions, assume that upon mutation the structural environment remains unaffected.

**The predictors**

The methods here developed were trained to predict whether a given single point protein mutation is classified in one of three classes: stabilising, destabilising and neutral. This task is addressed starting from the protein tertiary structure or from the protein sequence. For each task, the method is based on support vector machines (SVM) as implemented in libsvm release 2.7 (http://www.csie.ntu.edu.tw/~cjlin/libsvm/). We use a Radial Basis Functions kernel (RBF kernel = exp[-$G \| x_i - x_j \|^2$]). For the classification task we basically adopt the same input code by identifying three labels: one represents the increased protein stability ($\Delta\Delta G > 0.5$, label is +1), the second is associated with the destabilising mutation ($\Delta\Delta G < -0.5$, label is -1) and the last associated with neutral mutations ($-0.5 \leq \Delta\Delta G \leq 0.5$, label is 0). The input vector consists of 42 values. The first 2 input values account respectively for the temperature and the pH at which the stability of the mutated protein was experimentally determined. The next 20 values (for 20 residue types) explicitly define the mutation (we set to -1 the element corresponding to the deleted residue and to 1 the new residue (all the remaining elements are kept equal to 0). Finally, the last 20 input values encode the residue environment: namely a spatial environment, when the protein structure is available, or the nearest sequence neighbours, when only the protein sequence is available. When the protein structure is known (and the prediction is performed on the protein structure) each of the 20 values is the number of the encoded residue type, to be found inside a sphere of a 1.2 nm radius, centred on the coordinates of the C-alpha of the residue that undergoes mutation. Conversely, when the prediction is performed starting from the protein sequence, each of the 20 input values is again the number of the encoded residue type found inside a symmetrical window centred at the mutated residue, spanning the sequence towards the left (N-terminus) and the right (C-terminus), for a total length of 31 residues.

When prediction is structure-based, the Relative Solvent Accessible Area (RSA) value is calculated with the DSSP program [16], dividing the accessible surface area value of the mutated residue by the free residue surface. In this case a further input value (for a total sum of 43 numbers) includes the relative solvent accessible area of the mutated residue only when the protein structure is considered.

The input vectors associated to the reverse mutations are obtained by inverting the 20 values relative to the mutation elements and the others elements will be unchanged. The predictors here developed are compared with a SVM baseline algorithm that considers as input only the 20-element vector describing the residue mutation (SVM-BASE).

In order to compare the performance of our new three-state predictor with the previously developed method [13], we map the I-Mutant $\Delta\Delta G$ predicted values into the three defined classes, namely destabilizing mutations ($\Delta\Delta G < -0.5$ Kcal/mol), stabilizing mutations ($\Delta\Delta G > 0.5$ Kcal/mol) and neutral mutations ($-0.5 \leq \Delta\Delta G \leq 0.5$ Kcal/mol.).

**Scoring the performance**

The reported results on the different sets are evaluated using a 20 folds cross-validation procedure. The proteins considering in our datasets (DB3D and DBSEQ) are been clustered according to their sequence similarity using the *blastclust* program in the BLAST suite [17], by adopting the default value of length coverage equal to 0.9 and the score coverage threshold equal to 1.75. Furthermore, we keep the mutations that concern proteins in the same cluster and in the same position (when a residue is mutated in two different amino acids) in the same set, to minimise the possibility of an overestimation of the results. We also tested larger and a smaller partition of the database, but they do not significantly change the accuracy of our predictions. The data in the 20 sets used for cross validation are grouped in such a way that the stabilising and destabilising mutations are equally represented.

We also consider the new dataset (NewDB) and compare the performance of our method with the results derived form I-Mutant [13] predictions.

Several measures of accuracy are routinely used to evaluate machine learning based approaches. In this work we use the same measures of accuracy as previously reported [12,14], namely the overall accuracy (Q3), the sensitivity or coverage (Q), the specificity (P) and the correlation (C). In addition we also report the area ROC curve plotted calculating True Positive Rate TPR=TP/(TP+FN) and the False Positive Rate FPR=TP/(TP+FN), in order to show the distance from a random predictor (an area of 0.5 indicates random predictions).

# Results and Discussion

### Sequence-based Predictor

Previously we showed that it is possible to predict the sign of the $\Delta\Delta G$ using sequence and/or structure information [12-14]. Here, differently than before, we implement a SVM-based method that discriminates between stabilizing, destabilizing and neutral single point mutations. To optimize our method we consider different protein sequence contexts, and when starting from the sequence we analyse the effect of different lengths of the input window on the scoring efficiency (Table 1).

TABLE 1

It appears that the best scoring of our method is obtained when a window of 31 residues is taken into account, reaching an overall accuracy (Q3) of 0.52 and a mean correlation coefficient (<C>) of 0.28. The accuracy of our predictor is tested with respect to a baseline predictor that does not consider a sequence context (SVM-BASE). The sequence context improves the overall accuracy of 3% and the mean correlation of 2%. In Figure 2 we plot the overall accuracy and the mean correlation coefficient as a function of the reliability index (RI).

FIGURE 2

Noticing that the Q3 and <C> values increase at increasing values of the reliability index , we argue that the RI value may help in selecting which mutations are more suited to increase, decrease or leave unaltered the protein stability.

### Structure-based Predictor

The prediction of the sign and value of protein stability free energy change $\Delta\Delta G$ is more accurate when structural information is considered [12-14]. We implement this finding by considering spheres centred on the C-alpha of the mutated residues with different increasing radius values (see Table 2).

TABLE 2

In agreement with our previous work that considers an all heavy atom representation of the mutated residue, the best method for the three class discrimination is obtained when a radius of 12 Å is considered. The structure-based method reaches an overall accuracy of 0.58 (Q3) and a mean correlation coefficient (<C>) of 0.39. In order to provide a good indicator for selecting more reliable predictions, again Q3 and <C> values can be adopted given their increase as a function of the reliability index (RI) (Figure 3).

FIGURE 3

**Analysis of the prediction**
The sequence-based and the structure-based methods here proposed show a similar behaviour in the predictions of the three different classes of single point mutation. For the destabilizing ($\Delta\Delta G<-0.5$) and stabilizing ($\Delta\Delta G>0.5$) mutations obtained values of correlation coefficients are higher than those of neutral mutations (see Table 1 and 2). When the sequence and structural environments are considered, an improvement of the prediction of neutral mutations is detected. This is evident from the two different ROC curves of the stabilizing/destabilizing mutations (Figure 4A) compared to those of neutral mutations (Figure 4B). In the case of neutral mutations the increment of the ROC curve area is higher than that obtained when the baseline predictor is considered (Figure 4A).

Similar plots of the ROC curves are also reported for the structured-based method (see Figure 5). In this case higher values of ROC curve areas are generally obtained for all the three mutation classes and as before with sequence-based methods, the improving of the area for neutral mutations is greater that those obtained for stabilizing and destabilizing mutations (Figure 5).

FIGURE 5

When mutations with relevant effects on the protein stability ($|\Delta\Delta G|>0.5$ Kcal) are considered, the prediction of the destabilizing and stabilizing mutations is well balanced and reaches accuracy values of 71% and 76% with correlation coefficient of 0.43 and 0.52 for sequence-based and structure-based predictions, respectively.
Interestingly, the accuracy of our predictors can be evaluated as a function of the chemico-physical properties of the wild-type and of the mutated residues. The Q values obtained as a function of the chemical-physical type of wild type and mutated residue (from charged, polar and apolar to charged, polar and apolar residues, respectively) are shown for the sequence-based and structure-based methods, together with the abundance of the mutation type in the symmetric data base. Data are shown in Figures 6, 7 and 8 and reported with respect to destabilising, stabilising and neutral mutations, respectively. In the three different groups of mutations the most difficult to predict are those relative to the charged/charged and polar/charged residues. This is so irrespectively of the abundance in the symmetric data base (compare Figure 6, 7 and 8).

FIGURE 6

FIGURE 7

FIGURE 8

The general higher accuracy of the structure-based method with respect to the sequence-based ones is evident for each pair of mutations, and in agreement with what previously found [13]: it is more difficult to predict the protein stability change when mutations of charged/charged or polar/charged residues are considered (as indicated by lower mean correlation values, data not shown).

**Comparison between sequence-based and structure-based methods**

In order to better assess the quality of our predictors in relation with the provided output, we compare the prediction of sequence-based method with those obtained with the structure-based method. The comparison was performed selecting only the mutations associated to the proteins with known structure and dividing the DB3D dataset in three different range of relative accessible solvent area. In Figure 9 we report the overall accuracy (Q3) and the mean correlation coefficient <C> for highly buried residue (Relative Solvent Acces (RSA) ≤10%), for residues with 10%<RSA≤50% and exposed residue (RSA>50%).

Figure 9

We find that the larger differences between the sequence-based method SVM-WIN31 and the structure based-based method SVM-3D12 occur in the prediction of highly exposed residues, suggesting that when this is the case the structure-based code is better suited than that sequence-based to grasp the relevant features of the environment.

**Test and Comparison with previous methods**

We compare the new three-class discriminating implementation with our old two-class discriminating ones [13], by using a blind set: NewDB (see The protein data base section.).
In Table 3 the results of our two methods are compared with results obtained classifying the I-Mutant $\Delta\Delta G$ value output on the three different discriminated classes (as described in Material and Method). Even though the training data are the same, it is evident that the new SVM-based methods (SVM-WIN31 and SVM-3D12) achieve on average higher scores then the two algorithms of the previous I-Mutant predictor.

TABLE 3

More in details when sequence-based predictions are considered, the new method gains 3% in accuracy and 1% in correlation values; structure based predictions gain 6 % in accuracy and 5% in correlation.

# Conclusions

Our new development provides a more detailed prediction of the effects on the thermodynamics changes due to single point protein mutations considering that:
1) the thermodynamic reversibility adopted here generates a balanced data set that can help in over passing the problem of data-disproportion in favour of the large number of mutations associated to stability decrease found in the experimental data bases. Moreover the thermodynamic reversibility assumption makes the predictive methods intrinsically symmetric, similarly to the energy-based methods.

2) the introduction of a third class of neutral mutations grouping all the mutations that have a ΔΔG value close to 0 (-0.5 ≤ ΔΔG ≤ 0.5 Kcal/mol) partially prevents blurring in learning wrong associations due to the appreciable associated experimental errors.
We suggest that our new approach can be successfully applied when thermodynamic data of protein stability need to be analyzed in order to find more stabilizing/destabilizing mutations as compared to those that do not appreciably change the protein stability.

## Authors' contributions

EC contributes extracting data from ProTherm, implementing the predictors and writing the paper. PF, IR and RC contribute in the discussion of the thermodynamic hypothesis, in the review of the results and also in writing the paper.

## Acknowledgements

We thank MIUR for the following grants: PNR-2003 grant delivered to PF, a PNR 2001-2003 (FIRB art.8) and PNR 2003 projects (FIRB art.8) on Bioinformatics for Genomics and Proteomics and LIBI-Laboratorio Internazionale di BioInformatica, both delivered to RC. This work was also supported by the Biosapiens Network of Excellence project (a grant of the European Unions VI Framework Programme. EC acknowledge support from a Marie Curie International Reintegration Grant (FP6-039722) delivered to Marc A. Marti-Renom.

## References

1. Prevost M, Wodak SJ, Tidor B, Karplus M: **Contribution of the hydrophobic effect to protein stability: analysis based on simulations of the Ile-96-Ala mutation in barnase.** *Proc Natl Acad Sci U S A* 1991,. 88: 10880-10884.
2. Topham CM, Srinivasan N, Blundell TL: **Prediction of the stability of protein mutants based on structural environment-dependent amino acid substitution and propensity tables.** *Protein Eng.* 1997, 10: 7-21.
3. Pitera JW, Kollman PA: **Exhaustive mutagenesis in silico: multicoordinate free energy calculations on proteins and peptides.** *Proteins* 2000 41 :385-397.
4. Gilis D, Rooman M: **Predicting protein stability changes upon mutation using database-derived potentials: solvent accessibility determines the importance of local versus non-local interactions along the sequence.** *J. Mol. Biol.* 1997, 272: 276-290.
5. Kwasigroch JM, Gilis D, Dehouck Y, Rooman M: **PoPMuSiC, rationally designing point mutations in protein structures.** *Bioinformatics* 2002, 18: 1701-1702.
6. Funahashi J, Takano K, Yutani K: **Are the parameters of various stabilization factors estimated from mutant human lysozymes compatible with other proteins?** *Protein Eng.* 2001, 14: 127-134.

7. Guerois R, Nielsen JE, Serrano L: **Predicting changes in the stability of proteins and protein complexes: a study of more than 1000 mutations.** *J Mol Biol.* 2002, 320: 369-387.
8. Zhou H, Zhou Y: **Distance-scaled, finite ideal-gas reference state improves structure-derived potentials of mean force for structure selection and stability prediction.** *Protein Sci.* 2002, 11: 2714-2726
9. Parthiban V, Gromiha MM, Schomburg D: **CUPSAT: prediction of protein stability upon point mutations.** *Nucleic Acids Research* 2006, 34(Web Server issue):W239-42.
10. Parthiban V, Gromiha MM, Hoppe C, Schomburg D: **Structural analysis and prediction of protein mutant stability using distance and torsion potentials: role of secondary structure and solvent accessibility.** *Proteins* 2007, 66:41-52.
11. Casadio R, Compiani M, Fariselli P, Vivarelli F: **Predicting free energy contributions to the conformational stability of folded proteins from the residue sequence with radial basis function networks.** *Proc Int Conf Intell Syst Mol Biol.* 1995, 3: 81-88.
12. Capriotti E, Fariselli P, Casadio R: **A neural-network-based method for predicting protein stability changes upon single point mutations.** *Bioinformatics* 2004, 20 Supp 1: I63-I68.
13. Capriotti E, Fariselli P, Casadio R: **I-Mutant2.0: predicting stability changes upon mutatiotion from the protein sequence or structure.** *Nucleic Acids Research* 2005, 33 Web Server Issue: W306-W310.
14. Capriotti E, Fariselli P, Calabrese R, Casadio R: **Predicting protein stability changes from sequences using support vector machines.** *Bioinformatics* 2005, 21 Suppl 2: ii54-ii58.
15. Cheng J, Randall A, and Baldi P: **Prediction of protein stability changes for single-site mutations using support vector machines.** *Proteins* 2006, 62: 1125-1132.
16. Kumar MD, Bava KA, Gromiha MM, Prabakaran P, Kitajima K, Uedaira H, Sarai A: **ProTherm and ProNIT: thermodynamic databases for proteins and protein-nucleic acid interactions.** *Nucleic Acids Research* 2006, 34 (Database issue):D204-206.
17. Kabsch W, Sander C: **Dictionary of protein secondary structure: pattern of hydrogen-bonded and geometrical features.** *Biopolymers* 1983, 22: 2577-2637.
18. Altschul SF, Madden TL, Shaffer AA, Zhang J, Zhang Z, Miller W, Lipman DI: **Gapped BLAST and PSI-LAST a new generation of database search programs.** *Nucleic Acids Res.* 1997, 25: 3389-3402.

# Figures

**Figure 1 – Free Energy distribution of the database**

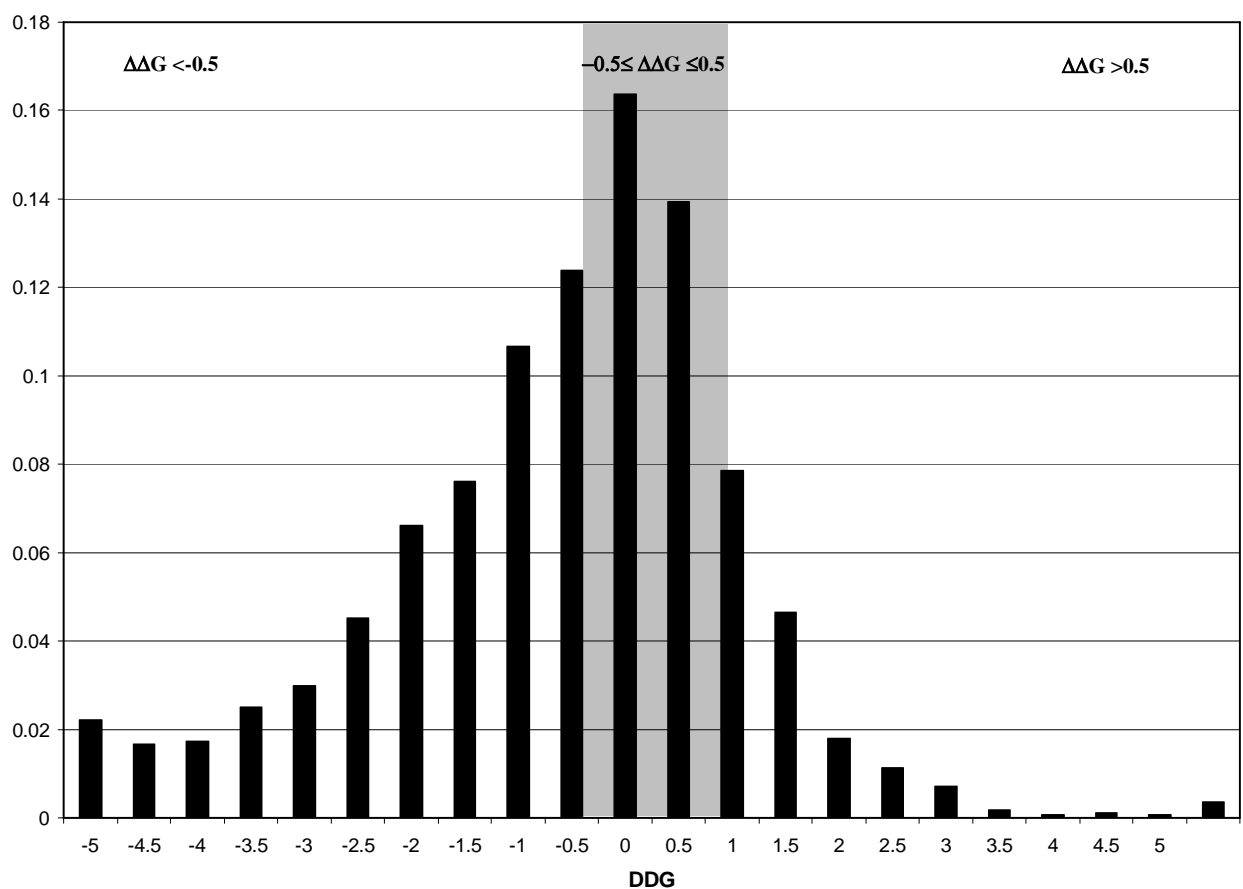

Distribution of DDG values on the 1681 mutations as extracted from the ProTherm database. The grey zone indicates neutral mutations in the data base.

**Figure 2 – Performances of the sequence-based predictor**

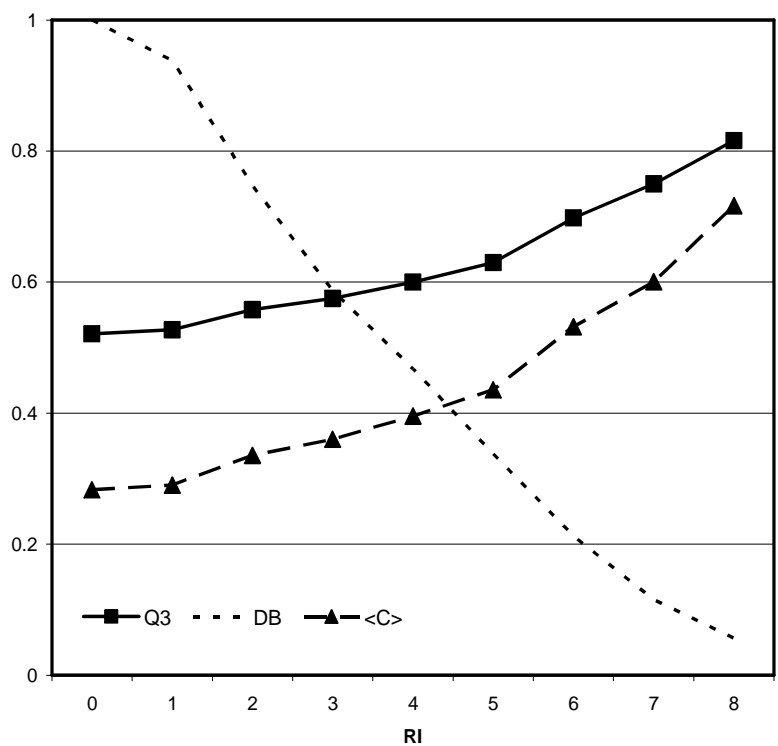

Overall accuracy (Q3) and correlation (C) of SVM-WIN31 as a function of the reliability index (RI) of the prediction. DB is the fraction of the data set DBSEQ with RI values higher or equal to a given threshold**.**

**Figure 3 – Performance of the structure-based predictor**

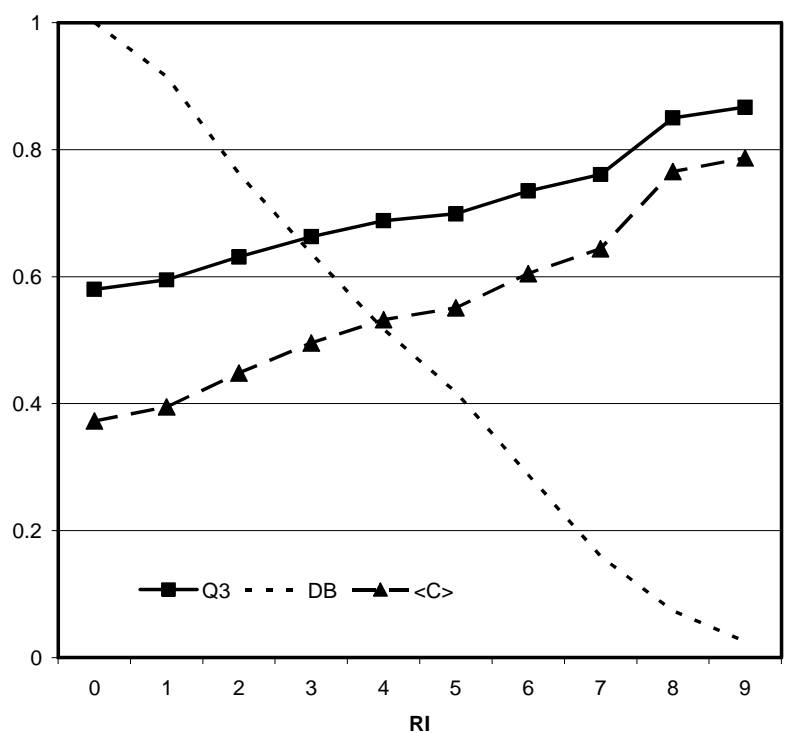

Overall accuracy (Q3) and correlation (C) of SVM-3D12 as a function of the reliability index (RI) of the prediction. DB is the fraction of the data set DB3D with RI values higher or equal to a given threshold.

**Figure 4 – ROC curves of the sequence-based predictor**

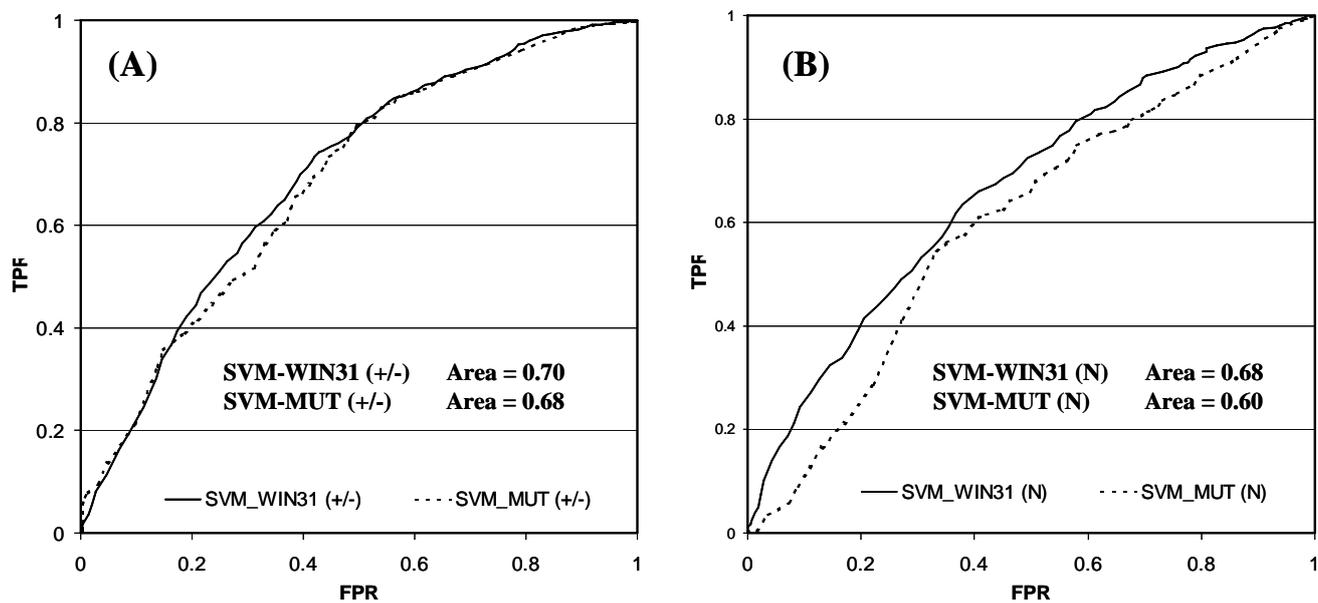

ROC curves for the sequence-based predictor. The cross-validation True Positive Rate (TPR) versus the False Positive Rate (FPR) are plotted the best method (SVM-WIN31) and for the baseline method (SVM-BASE). In part (A) the ROC curves of the two methods are relative to the prediction of increasing and decreasing free energy mutations ($|\Delta\Delta G|>0.5$ Kcal/mole), while in part (B) they are calculated for neutral mutations ($|\Delta\Delta G|\leq 0.5$ Kcal/mole).

**Figure 5 – ROC curves of the structure-based predictor**

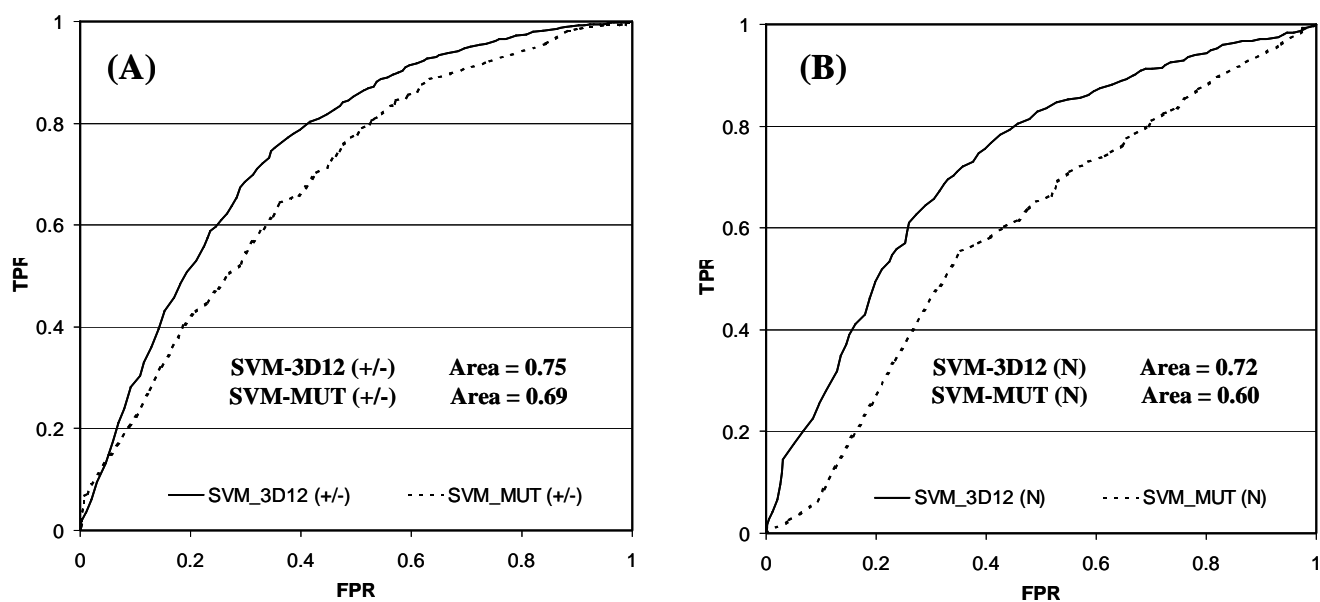

ROC curves for the structure-based predictor. The cross-validation True Positive Rate (TPR) versus the False Positive Rate (FPR) are plotted the best method (SVM-3DR12) and for the baseline method (SVM-BASE). In part (A) the ROC curves of the two methods are relative to the prediction of increasing and decreasing free energy mutations ($|\Delta\Delta G|>0.5$ Kcal/mole), while in part (B) they are calculated for neutral mutations ($|\Delta\Delta G|\leq 0.5$ Kcal/mole).

**Figure 6 – Analysis of the predictions on the destabilising mutations**

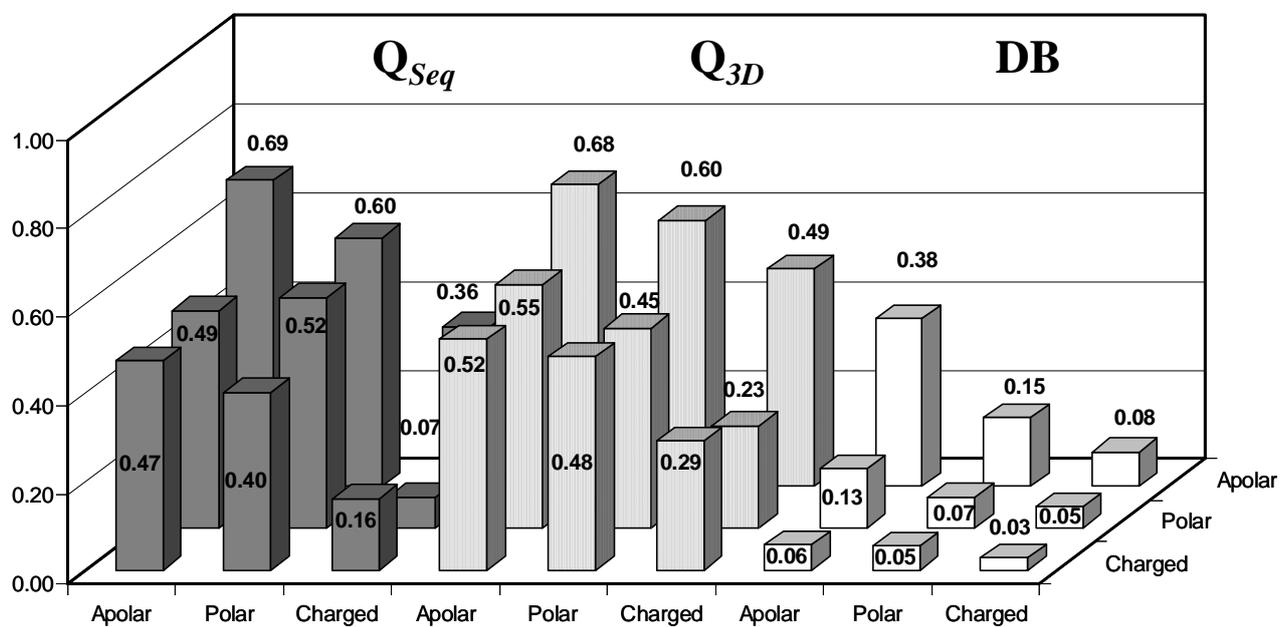

Accuracy [Q] of SVM-WIN31 (gray histograms), of SVM-3D12 (dotted histograms) and DBSEQ frequencies [DB] (white histograms) as a function of the mutated versus wild-type residues for the destabilizing mutations. The data are computed on the experimental database after symmetrising according the thermodynamic assumption (see Methods).

**Figure 7 – Analysis of the predictions on the stabilising mutations**

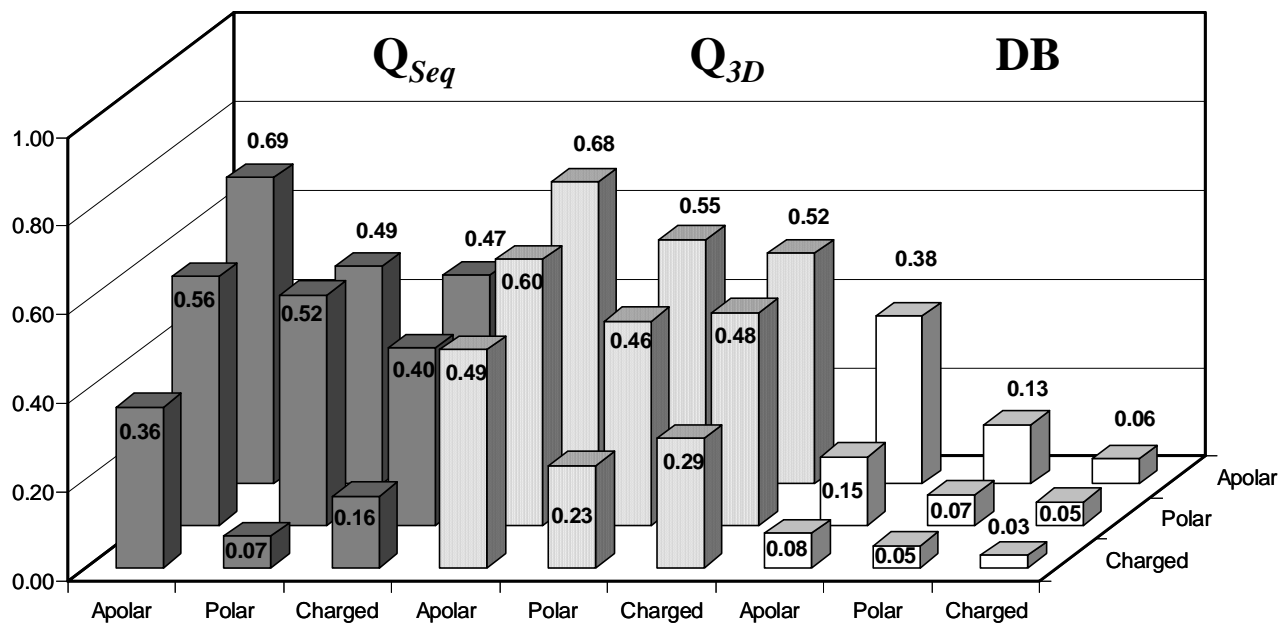

Accuracy [Q] of SVM-WIN31 (gray histograms), of SVM-3D12 (dotted histograms) and DBSEQ frequencies [DB] (white histograms) as a function of the mutated versus wild-type residues for the stabilizing mutations. The data are computed on the experimental database after symmetrising according the thermodynamic assumption (see Methods).

**Figure 8 – Analysis of the predictions on the neutral mutations**

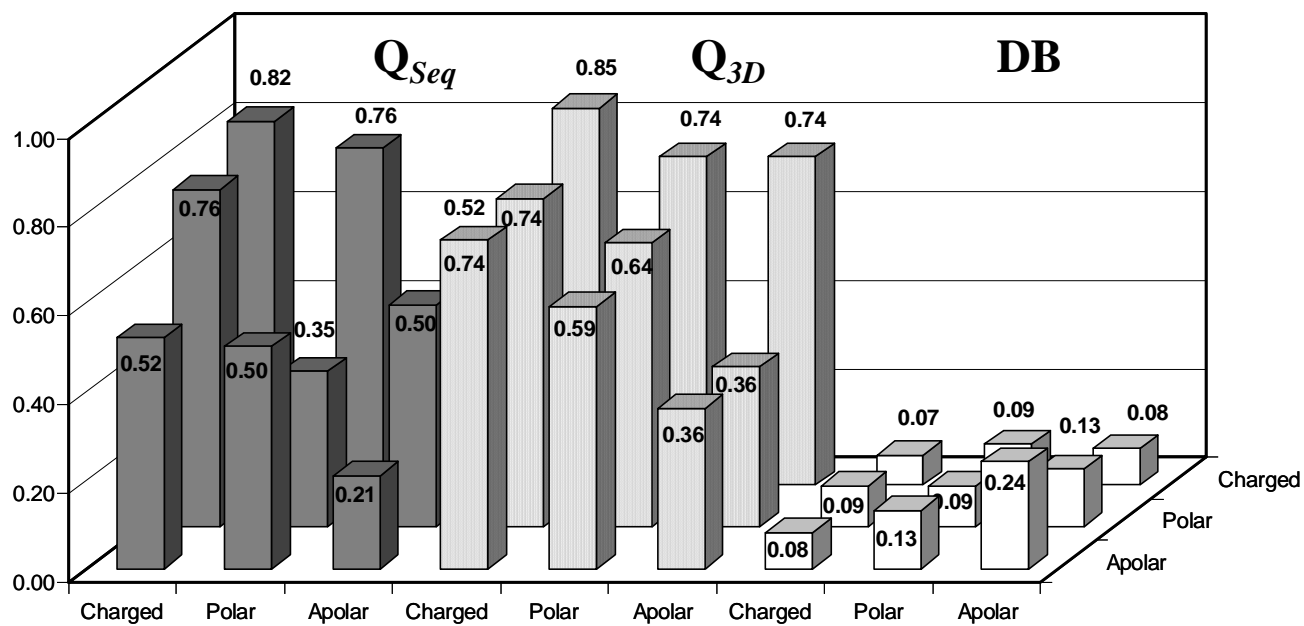

Accuracy [Q] of SVM-WIN31 (gray histograms), of SVM-3D12 (dotted histograms) and DBSEQ frequencies [DB] (white histograms) as a function of the mutated versus wild-type residues for the neutral mutations. The data are computed on the experimental database after symmetrising according the thermodynamic assumption (see Methods).

**Figure 9 – Comparison between sequence and structure base predictors**

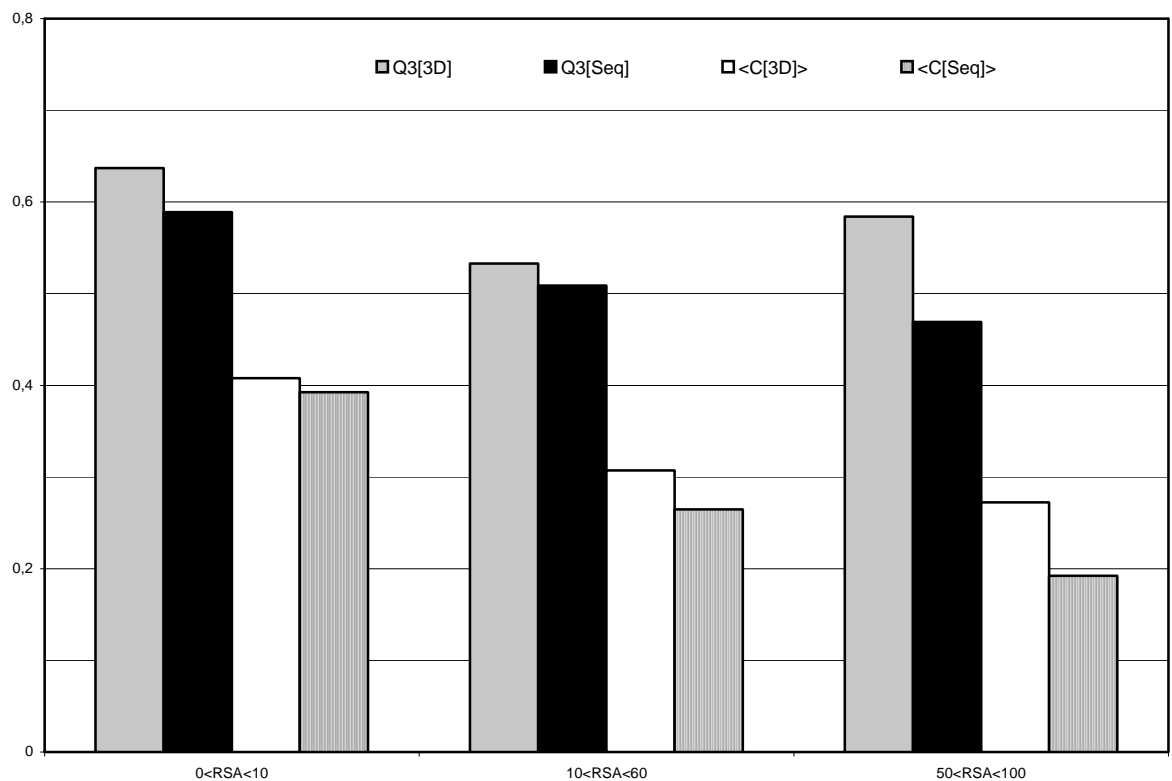

Overall accuracy (Q3) and mean correlation coefficient (<C>) of the structure-based (3D) and of the sequence-based (SEQ) methods calculated on mutations relative to highly buried residues with RSA≤10 (30% of DB3D), on mutations with 10<RSA≤50 (39% of DB3D) and on exposed residue RSA>50 (31% of DB3D).

# Tables

**Table 1 - Cross-validation performance of the sequence-based SVM method as a function of different windows lengths centred on the mutated residue**

| Method | Windows | Q3 | <C> | C[-] | P[-] | Q[-] | C[N] | P[N] | Q[N] | C[+] | P[+] | Q[+] |
|---|---|---|---|---|---|---|---|---|---|---|---|---|
| **SVM-BASE** | 0 | 0.49 | 0.24 | 0.28 | 0.52 | 0.54 | 0.15 | 0.43 | 0.40 | 0.28 | 0.52 | 0.54 |
| **SVM-WIN25** | 25 | 0.52 | 0.28 | 0.30 | 0.54 | 0.53 | 0.24 | 0.48 | 0.50 | 0.30 | 0.54 | 0.53 |
| **SVM-WIN31** | 31 | 0.52 | 0.28 | 0.31 | 0.55 | 0.53 | 0.24 | 0.48 | 0.50 | 0.30 | 0.54 | 0.53 |
| **SVM-WIN37** | 37 | 0.50 | 0.26 | 0.27 | 0.52 | 0.51 | 0.22 | 0.47 | 0.48 | 0.27 | 0.52 | 0.51 |

+ ,– and N: the index is evaluated for increasing , decreasing or neutral of protein free energy stability change, respectively according to the classification described in section 2; for the definition of the different indexes see the System and Methods in [12,14].

**Table 2 - Cross-validation performance of the structure-based SVM method as a function of different protein environments centred on the C-α of the mutated residue**

| Method | Radius (Å) | Q3 | <C> | C[-] | P[-] | Q[-] | C[N] | P[N] | Q[N] | C[+] | P[+] | Q[+] |
|---|---|---|---|---|---|---|---|---|---|---|---|---|
| **SVM-BASE** | 0 | 0.49 | 0.23 | 0.28 | 0.52 | 0.54 | 0.14 | 0.41 | 0.38 | 0.28 | 0.52 | 0.54 |
| **SVM-3D9** | 9 | 0.57 | 0.36 | 0.38 | 0.60 | 0.57 | 0.31 | 0.51 | 0.56 | 0.38 | 0.60 | 0.57 |
| **SVM-3D12** | 12 | 0.58 | 0.37 | 0.39 | 0.61 | 0.57 | 0.34 | 0.52 | 0.61 | 0.39 | 0.61 | 0.57 |
| **SVM-3D15** | 15 | 0.54 | 0.32 | 0.33 | 0.58 | 0.53 | 0.27 | 0.48 | 0.55 | 0.34 | 0.58 | 0.54 |

For notation see Table 1.

**Table 3 - Comparison of the performances of the best sequence-based SVM method (SVM-WIN31) and structure-based SVM method (SVM-3D12) with the I-Mutant based predictor.**

| Method | Q3 | <C> | C[-] | P[-] | Q[-] | C[N] | P[N] | Q[N] | C[+] | P[+] | Q[+] |
|---|---|---|---|---|---|---|---|---|---|---|---|
| **I-Mutant SEQ** | 0.49 | 0.27 | 0.37 | 0.48 | 0.86 | 0.15 | 0.43 | 0.39 | 0.30 | 0.79 | 0.22 |
| **SVM-WIN31** | 0.52 | 0.28 | 0.36 | 0.57 | 0.59 | 0.10 | 0.40 | 0.38 | 0.36 | 0.57 | 0.59 |
| **I-Mutant 3D** | 0.48 | 0.26 | 0.35 | 0.47 | 0.86 | 0.22 | 0.47 | 0.46 | 0.20 | 0.70 | 0.13 |
| **SVM-3D12** | 0.55 | 0.32 | 0.35 | 0.57 | 0.57 | 0.25 | 0.49 | 0.50 | 0.35 | 0.57 | 0.57 |

For notation see Table 1. I-Mutant SEQ, I-Mutant 3D, SVM-WIN3 and SVM-3D12 are tested on 998 mutations of NewDB database.

# Additional files

**Additional file 1 – DBSEQ**
The file containing the data used to train and testing the sequence based method is available at the http://lipid.biocomp.unibo.it/emidio/datasets/M3/DBSEQ_Sep05.txt .

**Additional file 2 – DB3D**
The file containing the data used to train and testing the structure based method is available at the http://lipid.biocomp.unibo.it/emidio/datasets/M3/DB3D_Sep05.txt .

**Additional file 2 – NewDB**
The file containing the data used to testing both the sequence and structure based method is available at the http://lipid.biocomp.unibo.it/emidio/datasets/M3/NewDB_Apr07.txt .